\documentclass[twocolumn,superscriptaddress,aps,prapplied, longbibliography]{revtex4-2} %Typical options above
\usepackage{graphicx} %insertion of eps figures
\usepackage{natbib} %implementation of \cite command; typical styles are apsrev is used for all the aps journals except rmp
\usepackage[usenames,dvipsnames]{color} % use colored text in latex
\usepackage{amsmath} % Defines extra environments for multiline displayed equations, as well as a number of other enhancements for math
\usepackage{amssymb} % {bm}% bold math
\usepackage{soul} %for underlining compatible with text wrapping; also other things
\usepackage{ifthen} %allows defining if then function
\usepackage{upgreek}
\usepackage[ampersand]{easylist}
\usepackage{todonotes}
\usepackage{verbatim}

\newcommand{\appsection}[1]{\section{\MakeUppercase{#1}}}

\newcommand{\bra}[1]{\langle{#1}|}
\newcommand{\ket}[1]{|{#1}\rangle}

\def\be#1\ee{\begin{equation}#1\end{equation}}
\def\ba#1\ea{\begin{align}#1\end{align}}
\def\bg#1\eg{\begin{gather}#1\end{gather}}
\def\t{\text}

\newcommand{\vect}[1]{\boldsymbol{#1}}
\newcommand{\abs}[1]{\lvert#1\rvert}

\begin{document}

%\detailtexcount{AnharmonicCSFQ_v2}
%TC:ignore
\title{Characterization of multi-level dynamics and decoherence in a high-anharmonicity capacitively shunted flux circuit}

\author{M.A. Yurtalan}
\email{mayurtalan@uwaterloo.ca}
\affiliation{Institute for Quantum Computing, University
	of Waterloo, Waterloo, ON, Canada N2L 3G1}
\affiliation{Department of Physics and Astronomy,  University
	of Waterloo, Waterloo, ON, Canada N2L 3G1}
\affiliation{Department of Electrical and Computer Engineering,  University
	of Waterloo, Waterloo, ON, Canada N2L 3G1}

\author{J. Shi}
\affiliation{Institute for Quantum Computing, University
	of Waterloo, Waterloo, ON, Canada N2L 3G1}
\affiliation{Department of Physics and Astronomy,  University
	of Waterloo, Waterloo, ON, Canada N2L 3G1}

\author{G.J.K. Flatt}
\affiliation{Department of Physics and Astronomy, University of California, Los Angeles, Los Angeles, CA, United States 90095}

\author{A. Lupascu}
\email{alupascu@uwaterloo.ca}
\affiliation{Institute for Quantum Computing, University
	of Waterloo, Waterloo, ON, Canada N2L 3G1}
\affiliation{Department of Physics and Astronomy,  University
	of Waterloo, Waterloo, ON, Canada N2L 3G1}
\affiliation{Waterloo Institute for Nanotechnology, University
of Waterloo, Waterloo, ON, Canada N2L 3G1}

%TC:endignore

\date{ \today}

%TC:ignore
\begin{abstract}%%%****************************************************ABSTRACT

We present the design and characterization of a three-Josephson-junction superconducting loop circuit with three large shunt capacitors. The circuit used as a qubit shows long energy relaxation times, of the order of 40 $\upmu$s, and a spin-echo dephasing time of 9.4 $\upmu$s. The circuit has high anharmonicity, of $2\pi\times3.69$ GHz. We extract the multi-level relaxation and dephasing rates of the circuit used as a qutrit and discuss the  possible sources for the decoherence. The high anharmonicity allows for fast qubit control with nanosecond range gate durations and a measured average gate fidelity of  99.92\%, characterized by randomized benchmarking. These results demonstrate interesting potential use for fast nanosecond time scale two-qubit gates and multi-level quantum logic.
\end{abstract}
%TC:endignore

\maketitle
\section{Introduction}
Recent years were marked by major progress in gate-model quantum computing implementations, leading to the development of small prototypes with sizes reaching tens of qubits~\cite{arute_quantum_2019,corcoles2019challenges,wright2019benchmarking,otterbach_unsupervised_2017}. Superconducting quantum bits in particular received attention as one of the most promising platforms from the perspective of scalability~\cite{kelly_state_2015,gambetta_building_2017,otterbach_unsupervised_2017}. Despite these advances, research on improved quantum bits and on methods for implementation of elementary single- and two-qubit gates remains a highly relevant research topic. Single- and two-qubit gate fidelities have only approached or marginally exceeded the error tolerance threshold for the surface code~\cite{barends_superconducting_2014, chow_universal_2012}. Reducing gate errors has the potential to lead to a dramatic reduction in fault tolerant operation overhead~\cite{fowler_surface_2012} and is relevant for non-error-corrected near-term quantum devices~\cite{boixo_characterizing_2018}. Gate errors are determined by both qubit coherence times and gate speed, and more generally by architecture details; this complete design space has been only partially explored in superconducting devices.

In this paper, we present experimental results on a superconducting qubit design that combines long coherence times, of $40\,\upmu$s for energy relaxation and $9.4\,\upmu$s for spin-echo dephasing, with high level anharmonicity of $2\pi\times3.69$~GHz. Level anharmonicity is the difference between the first two transition frequencies, $\omega_{12}$ and $\omega_{01}$, with $0$ and $1$ the first two energy eigenstates, used as qubit computational states, and $2$ the second excited state. Anharmonicity has a direct impact on the speed of single-qubit gates~\cite{motzoi_simple_2009,chow_optimized_2010} and is a limiting factor for the speed of two-qubit gate implementations~\cite{groot_selective_2012}. We demonstrate fast single-qubit gates, with duration of 1.62 ns for a $\pi/2$ pulse and 2.64 ns for a $\pi$ pulse,  and a high fidelity, characterized using randomized benchmarking, reaching $99.92\,\%$. The large anharmonicity of this design, combined with the long coherence times, have the potential to lead in the future to fast and high-fidelity two-qubit gates. Moreover, we performed experiments in which we characterized decoherence and control of this device used as a qutrit (formed by states $0$, $1$, and $2$). Qutrit control and coherence bear relevance for qubit gates that make use of the properties of higher levels and is more broadly relevant for quantum protocols based on multi-level logic~\cite{campbell_enhanced_2014,gottesman_fault-tolerant_1999,muthukrishnan_multivalued_2000}. These results are enabled by a flux-type qubit design with three junctions with large planar capacitive shunts. Previous work on capacitively shunted flux qubit circuits focused on single shunt designs and the dynamics and properties of the lowest two levels. Capacitive shunting of flux qubits was proposed as an approach to reduce charge noise induced decoherence~\cite{you_low-decoherence_2007, steffen_high-coherence_2010}. More recently Yan \emph{et al.}~\cite{yan_flux_2016} performed a systematic study of flux qubits with single capacitive shunts, and demonstrated high coherence and a moderate anharmonicity approaching 1~GHz in optimized samples. Stern \emph{et al.}~\cite{stern_flux_2014} demonstrated relatively long energy relaxation times in flux qubits coupled to three-dimensional cavities, suitable for hybrid device experiments.

\begin{figure}[h]
\centering
\includegraphics[width=3.4in]{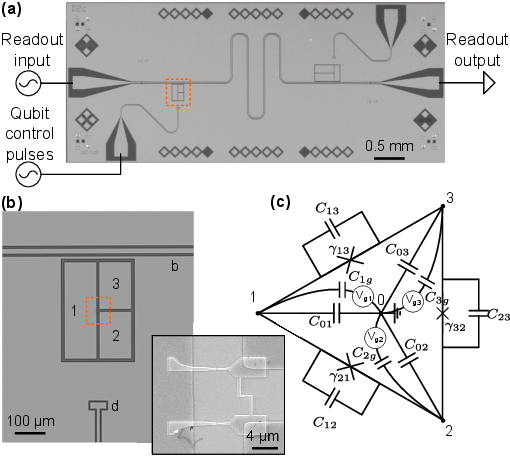}
\caption{\label{fig:device} (Color online) (a) A photograph of the capacitively-shunted flux qubit coupled to a CPW resonator. (b) A microscope image, corresponding to the region indicated by dashed rectangle in (a), showing the shunt capacitor pads, resonator electrode, and control electrode (denoted by 1, 2, 3, b, and d, respectively). The inset shows scanning electron image of the qubit loop and junctions placed in between the capacitor pads, corresponding to the region indicated by dashed rectangle in (b).(c) The circuit diagram of the device. The nodes correspond to the electrodes shown in (b).}
\end{figure}

\section{Design and Circuit Model}
We developed a flux qubit design with three Josephson junctions and three capacitive pads (see Fig.~\ref{fig:device}).  Two junctions are of equal size and the third is smaller by a factor $\alpha=0.61$. The use of  large capacitive pads is intended to effectively reduce the participation ratio of electric fields in Josephson junctions and at metal surfaces and interfaces, as demonstrated in transmons~\cite{koch_charge-insensitive_2007} and then adapted to other types of superconducting qubits~\cite{steffen_high-coherence_2010,yan_flux_2016}. In our design, the use of three pads gives additional flexibility in the design (see Appendix~\ref{sec:appopt}), allowing to reach simultaneously favorable metrics for control (large anharmonicity) and decoherence (low flux noise and charge noise sensitivity). The qubit is biased with a magnetic flux $\Phi$ generated by an external coil. Transmission through the resonator is used to measure the state of the qubit, based on dispersive-regime circuit quantum electrodynamics~\cite{wallraff_approaching_2005}. A coplanar waveguide (CPW) terminated by a capacitive pad is used to drive transitions between the circuit energy eigenstates. The qubit is placed inside a sample holder at the mixing chamber of a dilution refrigerator. All transmission lines contain attenuators, low-pass filters, and infrared filters.

%\section{circuit model}
The circuit model for the capacitively shunted three-Josephson junction loop device is shown in Fig.~\ref{fig:device}. The model includes the capacitances between nodes 0, 1, 2, and 3, which correspond, respectively, to the ground plane and to the three shunt capacitor pads. The  resonator electrode and the control electrode are capacitively coupled to the pads and modeled by the capacitances between the electrodes and the pads with $\mathbf{C}_\t{b}=(C_{1\t{b}},C_{2\t{b}},C_{3\t{b}})$  and $\mathbf{C}_\t{d}=(C_{1\t{d}},C_{2\t{d}},C_{3\t{d}})$, together with  the voltages $V_\t{b}$ and $V_\t{d}$, respectively. Trapped charges are modelled  by gate capacitances and gate voltages and denoted by the charge vector $\mathbf{Q}_\t{g}$. The superconducting phase differences across the junctions are denoted by $\gamma_{21}$, $\gamma_{31}$, and $\gamma_{23}$ and the branch phase between nodes 0 and 1 is represented by $\gamma_{01}$. The Hamiltonian of the circuit is given by
\begin{equation}
H=T_\t{H}+U_\t{H},
\label{eq:ham}
\end{equation}
with the kinetic energy term
\begin{eqnarray}
T_\t{H}&=&\frac{1}{2\varphi_0} \big [ \mathbf{p}  - \varphi_0 \mathbf{D} ( {V}_\t{b}\mathbf{C}_\t{b} + {V}_\t{d}\mathbf{C}_\t{d} + \mathbf{Q}_\t{g}  )  \big ] \nonumber \\ && \mathbf{C}^{-1} \big [ \mathbf{p}  - \varphi_0 \mathbf{D} ( {V}_\t{b}\mathbf{C}_\t{b} + {V}_\t{d}\mathbf{C}_\t{d} + \mathbf{Q}_\t{g}  )   \big ]^T,\label{eq:hamt}
\end{eqnarray}
and the potential energy term
\begin{eqnarray}
U_\t{H}&=&-\varphi_0 I_\t{c}[  \cos(\gamma_{21}) + \cos(\gamma_{31}) \nonumber \\  &&+\alpha \cos(\gamma_{21}-\gamma_{31}+2\pi \frac{\Phi}{\Phi_0}) ],
\end{eqnarray}  
where $\varphi_0=\Phi_0/2\pi$ is the reduced magnetic flux quantum, $\Phi$ is the external magnetic flux,  $I_\t{c}$ is the junction critical current of the larger junctions, and $\mathbf{p}=(p_{\gamma_{21}},p_{\gamma_{31}},p_{\gamma_{01}})$ is the  vector of the momenta conjugate to the corresponding phases. The capacitance matrix  of the system $\mathbf{C}$ is given by
\begin{equation}
\mathbf{C} =
\begin{pmatrix}
\begin{smallmatrix}  
\substack{C_{12}+C_{23}+C_\t{2g}\\+C_{02}+C_\t{2b}+C_\t{2d} \\ \ }  &-C_{23} 			  & \substack{C_\t{2g}+C_{02}\\+C_\t{2b}+C_\t{2d}}\\ \\
-C_{23}		           & \substack{C_{13}+C_{23}+C_\t{3g}\\+C_{03}+C_\t{3b}+C_\t{3d} }  & \substack{C_\t{3g}+C_{03}\\ + C_\t{3b} +C_\t{3d} \\ \ }\\ \\
\substack{C_\t{2g}+C_{02}\\+C_\t{2b}+C_\t{2d}}    & \substack{C_\t{3g}+C_{03}\\+C_\t{3b} +C_\t{3d}}	  & \substack{C_\t{1g}+C_\t{2g}+C_\t{3g} \\ +C_{01}+C_{02}+C_{01} \\ +C_\t{1b}+C_\t{2b}+C_\t{3b}  \\  +C_\t{1d}+C_\t{2d}+C_\t{3d}} 
\end{smallmatrix}
\end{pmatrix}.
\label{eq:cmatrix}
\end{equation}
The gate capacitances for trapped charges $C_\t{1g}$, $C_\t{2g}$, and $C_\t{3g}$ are assumed to be negligibly small. The matrix $\mathbf{D}$ is given by 
\be
\mathbf{D}=
\begin{pmatrix}
	0 &-1 & 0 \\
	0 & 0 &-1 \\
	-1&-1 &-1
\end{pmatrix}.
\ee
The capacitance values in the system capacitance matrix in Eq.~\ref{eq:cmatrix} are numerically determined  using finite-element electromagnetic simulation tools, and are given in Table~\ref{tbl:simcaps}.

\begin{table}[h]
	\centering
	\caption{The  capacitances of the  device geometry. }
	\begin{tabular*}{3.4in}{c@{\extracolsep{\fill}}c@{\extracolsep{\fill}}c@{\extracolsep{\fill}}c}
		\hline
		\hline
		Capacitor & Value & Capacitor & Value\\
		\hline
		$C_\t{13}$& $17.91 \t{ fF}$ & $C_\t{21}$&$ 18.13 \t{ fF}$ \\  $C_\t{32}$& $10.56 \t{ fF}$  & $C_\t{01}$&$62.9 \t{ fF}$ \\ $C_\t{02}$& $30.4 \t{ fF}$ & $C_\t{03}$&$ 32.9 \t{ fF}$  \\
		$C_\t{1b}$& $2.60 \t{ fF}$ & $C_\t{2b}$&$ 2.61 \t{ fF}$ \\ $C_\t{3b}$& $0.21 \t{ fF}$ & $C_\t{1d}$&$ 0.15 \t{ fF}$ \\ $C_\t{2d}$& $0.02 \t{ fF}$ & $C_\t{3d}$&$ 0.11 \t{ fF}$ \\
		\hline
		\hline
	\end{tabular*}
	\label{tbl:simcaps}
\end{table}

\section{experimental techniques}

In this section, we present the methods and techniques for the analysis of the device characteristics. First, we discuss the interplay of energy relaxation and pure dephasing in multi-level decoherence. Second, we detail the noise spectroscopy method  with  Carr-Purcell-Meiboom-Gill (CPMG) experiments for $1/\abs{\omega}^\delta$ type noise. Last, we discuss dephasing due to photon noise in the readout resonator.

\subsection{Multi-level Dephasing}
For a two-level system, a Ramsey experiment is used to characterize the decay of the off-diagonal matrix element $\rho_{01}$ of the density matrix. In a frame resonant with the transition frequency, this decay is given by $\rho_{01}(t) = \mathcal{C}(t) e^{-t/(2T_1)}$, where $T_1$ is the energy relaxation time and $\mathcal{C}(t)$ is the coherence function. If noise is modelled as a classical stochastic process $\xi(t)$ contributing a term to the Hamiltonian
$H_\t{qb,r}=-\hbar\xi(t)\vect\sigma_z/2$,
which is diagonal in the energy eigenbasis, the coherence function is given by
\begin{equation}
\mathcal{C}(t)=  \left < \exp\left(-i\int_0^t\xi(t')dt'\right) \right >, 
\end{equation}
where $\langle .. \rangle$ is an average over noise realizations.

In this section we discuss the generalization pure dephasing decay to a $n$-level system. We assume the noise couples diagonally, as a term to the system Hamiltonian
\begin{equation}
\label{eq:HrMultiLevel}
H_\t{r}(t)=-\hbar\sum_{j=0}^{n-1}\xi_j(t)\ket{j}\bra{j},
\end{equation}
where $\ket{j}$, $j=\overline{0,n-1}$, are the energy eigenstates and $\xi_j(t)$ are random noise processes. We analyze the dynamics of the multi-level system under simultaneous coupling to a Markovian bath and coupling to a noise source described by Eq.~\ref{eq:HrMultiLevel}. It can be shown that in a rotating frame given by the qubit nominal Hamiltonian, the evolution of the off-diagonal terms of the density matrix $\rho_{jk}(t)$ is given by
\begin{equation}
\rho_{jk}(t)=\exp \left (i\int_0^t[\xi_j(t')-\xi_k(t')]\,dt'\right)\tilde\rho_{jk}(t),
\end{equation}
where $\vect{\tilde\rho}(t)$ is the density matrix in the rotating frame in the absence of noise terms ($H_\t{r}(t) = 0$). When averaging is done over different realizations of the noise, we obtain
\begin{equation}
\label{eq:OffDiagonalRhoEvolution}
\rho_{jk}(t)=\mathcal{C}_{jk}(t) \tilde\rho_{jk}(t),
\end{equation}
where the generalized coherence function is
\begin{equation}
\mathcal{C}_{jk}(t) =  \left < \exp\left (i\int_0^t[\xi_j(t')-\xi_k(t')]\,dt'\right) \right >.
\end{equation}

Formally, we can write the transformation of the density matrix in the rotating frame from initial time $t_i$ to final time $t_f$ as 
\begin{equation}
\label{eq:CombinedRelaxationAndDephasing}
\vect\rho(t_f) = \mathcal{D}[\mathcal{R}[\vect\rho(t_i)]],
\end{equation}
where $\mathcal{R}$ is an operator that describes Markovian relaxation and $\mathcal{D}$ is an operator that acts on the off-diagonal elements of the density matrix  according to Eq.~\ref{eq:OffDiagonalRhoEvolution}. This equation is most conveniently expressed by using the density matrix in column form (see e.g.~Ref~\cite{havel_RobustProceduresConverting_2003}). To characterize multi-level dephasing in experiments, we proceed as follows. First, multi-level relaxation is characterized in an experiment where we prepare an excited state and we measure the decay of populations versus time, as discussed in the previous subsection. Next, we prepare a superposition of states $\ket{j}$ and $\ket{k}$, and monitor the decay of $\rho_{jk}$ in a Ramsey type experiment. After factoring out energy relaxation terms based on Eq.~\ref{eq:CombinedRelaxationAndDephasing}, the coherence function $\mathcal{C}_{jk}(t)$ is obtained.

\subsection{Decoherence with $A/\abs{\omega}^\delta$ noise}
\label{sec:cpmg2}
In this section, we discuss qubit coherence measurements using CPMG sequences~\cite{bylander_NoiseSpectroscopyDynamical_2011,orgiazzi_flux_2016}. With Gaussian noise, the coherence function for a CPMG sequence with N pulses is given by
\begin{equation}
\label{eq:CPMGCoherenceDecay}
    \mathcal{C}_N(\tau) =\exp{\bigg [     \int_{-\infty}^{\infty}d\omega\, S(\omega)F(\omega,N,\tau)          \bigg ]},
\end{equation}
where $S(\omega)$ is the double sided noise power spectral density (PSD) of fluctuations in qubit angular transition frequency. The filter function $F(\omega,N,\tau)$ is given by 
\begin{equation}
    F(\omega,N,\tau)=\frac{1}{2} \bigg | {\int_{0}^{\tau}dt\, \zeta(N,t)e^{i\omega t}} \bigg|^2,
\label{eq:filter1}    
\end{equation}
where $\zeta(N,t)\in \{-1,1\}$ is the CPMG sign multiplier for the noise after each refocusing $\pi$  rotation. Using the dimensionless parameter $X=\omega\tau$ and defining $\bar{F}(X,N)= F(X,N,\tau)/\tau^2$, we find
% \setstretch{1.5}
\begin{equation}
\bar{F}(X,N)=
\begin{cases}
\frac{8}{X^2}\sin^4 (\frac{X}{4N})\frac{\cos^2({X}/{2})}{\cos^2({X}/{2N})},&\text{if } N \text{ is odd, } \\[5pt]
\frac{8}{X^2}\sin^4 (\frac{X}{4N})\frac{\sin^2({X}/{2})}{\cos^2({X}/{2N})},&\text{if } N \text{ is even. }\\
\end{cases}
\end{equation}
We note that $\bar{F(X,N)}$  is peaked at $X\approx N \pi$, with a peak width of the order 1. We assume we can neglect the variations of the noise PSD $S(\omega)$ over the peak of the filter function. Defining the integrals $I =\int dX\, \bar{F}(X,N,\tau)$ and $X^*=(1/\t{I})\int dX\,  \bar{F}(X,N,\tau) X$, the coherence function can be approximated as
\begin{equation}
\label{eq:CPMGCoherenceDecayApprox}
    \mathcal{C}_N(\tau) \approx \exp\bigg [-2\tau \t{I} S\big (\frac{X^*}{\tau}\big )\bigg].
\end{equation}
We note that $I \approx1.24$ and $X^*\approx\pi\times N$ are good approximations for $1<N<200$. Eq.~\ref{eq:CPMGCoherenceDecayApprox} allows for finding the noise PSD directly from the measurement of the coherence function. We next discuss the case when  $S(\omega)=A/\abs{\omega}^\delta$. In this case Eq.~\ref{eq:CPMGCoherenceDecayApprox} becomes
\begin{equation}
 \mathcal{C}_N(\tau)=\exp{\big[ -(\Gamma_N \tau)^{(\delta+1)}   \big]},
 \label{eq:coh3}
\end{equation}
with
\begin{equation}
\Gamma_N=(2.48A)^{1/(\delta+1)}(\pi N)^{(-\delta/\delta+1)}.
\label{eq:coh4}
\end{equation}
\subsection{Decoherence due to photon noise}
We consider dephasing of the qubit at the symmetry point due to fluctuations of the photon number in the cavity. We use numerical simulations to predict dephasing due to this source. Specifically, the photon population of the cavity is modeled using a random telegraph noise with states $n=0$ (empty cavity) and $n=1$ (cavity occupied by one photon). The transition rates between two states are $\Gamma_{0 \rightarrow 1} = \omega_r / Q \times n_\t{th}$ and $\Gamma_{1 \rightarrow 0} = \omega_r / Q \times  (1 + n_\t{th})$, with $\omega_r$ the cavity resonance frequency, $Q$ the cavity quality factor, and $n_\t{th}$ the thermal photon number~\cite{haroche_ExploringQuantumAtoms_2006}. The effect of photon number fluctuations on the qubit is determined by the dispersive shift. This numerical approach is then compared with the analytical expression for Ramsey decay in Ref.~\cite{rigetti_SuperconductingQubitWaveguide_2012}.

\section{Experimental results}
\begin{figure}[h]
\centering
\includegraphics[width=3.4in]{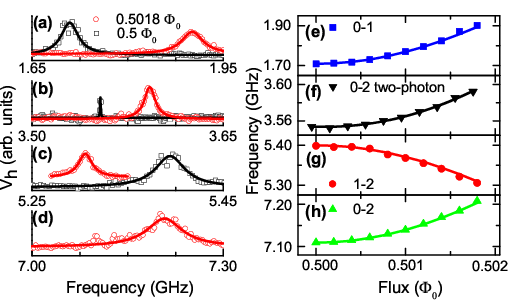}
\caption{\label{fig:spectroscopy}(Color online) (a-d) Average homodyne voltage versus the frequency of the applied spectroscopy pulse, showing (a) 0-1, (b) 0-2 two-photon, (c) 1-2, and (d) 0-2 transitions, respectively  at  $\Phi=0.5 \,\Phi_0$ (black squares) and  $\Phi=0.5018 \,\Phi_0$ (red circles). Solid lines represent Lorentzian fits. (e-h) The transition frequency versus the applied external flux $\Phi$ for (e) 0-1, (f) 0-2 two-photon, (g) 1-2, and (h) 0-2  transitions. Solid lines are fits of the transition frequency based on the circuit model.  }
\end{figure}

We first present the results of the spectroscopy experiments. In Fig.~\ref{fig:spectroscopy}(a)-(d) we show the readout result, given by the average homodyne voltage, versus the frequency of the applied spectroscopy pulse, for two values of the applied magnetic flux - $\Phi=0.5~\Phi_0$ (the flux symmetry point) and $\Phi=0.5018~\Phi_0$.  At the symmetry point, we observe a  peak at the transition frequency $\omega_{01}=2\pi\times 1.708~\t{GHz}$ between states 0 and 1 and a peak at $\omega_{02}^\t{tp}=2\pi\times 3.553~\t{GHz}$, a two-photon transition between states 0 and 2.  At $\Phi=0.5018~\Phi_0$, we observe the 0-1 and 0-2 two-photon transitions as well as the 0-2 transition, with the latter absent at the symmetry point due to selection rules~\cite{liu_optical_2005}. The 1-2 transition, with a frequency of $\omega_{12}=2\pi\times 5.398~\t{GHz}$ at the symmetry point, can be observed after applying a $01$ pumping tone. The  transition frequencies $\omega_{01}$, $\omega_{02}^\t{tp}$, $\omega_{12}$, $\omega_{02}$, for a range of applied magnetic flux, are shown in Fig.~\ref{fig:spectroscopy}(e)-(h). 
The numerically calculated energy levels, extracted based on the circuit Hamiltonian Eq.~\ref{eq:ham}, are used to fit the experimentally measured first two transition frequencies versus applied flux. The capacitance values listed in  Table~\ref{tbl:simcaps} are used as fixed parameters and the  junction critical currents are used as the only adjustable fit parameters. We observe an excellent agreement between the fitted data and the experiments with the best fit corresponding to a junction critical current density of $3.96$ $\upmu\t{A}/\upmu\t{m}^2$  and $\alpha=0.61$, which are in good agreement with the target nominal critical current density of $4.0$ $\upmu\t{A}/\upmu\t{m}^2$ and the design value of $0.62$ for $\alpha$. For the smaller junction, the ratio of the Josephson energy $E_\t{J} = \Phi_0 I_c / (2 \pi)$ to the charging energy $E_\t{C} = (2e)^2/(2 C_{23})$  is $E_\t{J} / E_\t{C} = 6.9$, with $E_\t{J}/h = 50.9$ GHz.
\begin{figure}[h]
\centering
\includegraphics[width=3.4in]{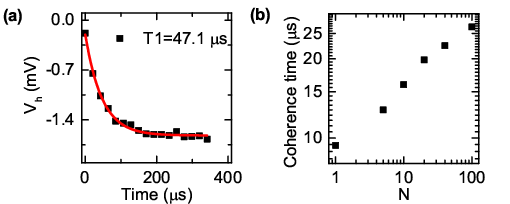}
\caption{\label{fig:coherencesym}(Color online) (a) Readout voltage versus readout delay time after a $\pi^{01}$ pulse is applied to the qubit thermalized state. The solid line is an exponential fit, leading to a relaxation time $T_1 = 47.1 \upmu$s. (b) Coherence time versus the number of pulses N in a CPMG sequence.  
}
\end{figure}

We next discuss the qubit coherence characterization at the symmetry point ($\Phi=0.5~\Phi_0$). Fig.~\ref{fig:coherencesym}(a) shows an energy relaxation time measurement, with the qubit excited with a $\pi_x^{01}$ rotation, resonant with the 0-1 transition. The measured $T_1$ had systematic fluctuations in the course of experiments, with most values between 35 and 45 $\upmu$s and reached values as high as $T_1=47.1 \pm 2.0 \,\upmu$s. The measured relaxation times are comparable to the best result reported previously on a capacitively shunted flux qubit design with moderate anharmonicity~\cite{yan_flux_2016}. 

The Ramsey coherence times, extracted with exponential and Gaussian decaying fits, were found to be $4.7$ and 7.3 $\upmu$s, respectively.  For Ramsey dephasing, exponential decay as opposed to Gaussian decay led to a better fit. The spin-echo coherence time was found to be T$_\t{2E}=9.4$ $\upmu$s.  We performed dephasing measurements with dynamical decoupling based on CPMG pulse sequences with $\pi$ pulse duration of 11.9 ns. The coherence time versus the number of pulses is shown in Fig.~\ref{fig:coherencesym}(b). The coherence time reaches $T_\t{CPMG} = 26.5\,\upmu$s for $N=100$ pulses. Gaussian fits were used to extract the spin echo and CPMG decay times.

\begin{figure}[h]
\centering
\includegraphics[width=3.4in]{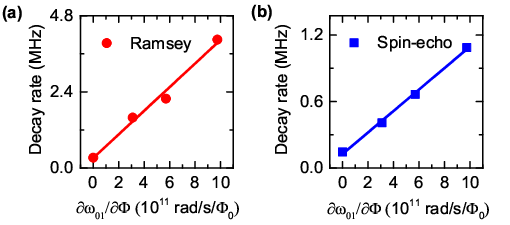}
\caption{\label{fig:fig5}(a) Ramsey and (b) spin-echo decay rates versus the flux sensitivity coefficient.  }
\end{figure}

In addition to the experiments at the flux symmetry point, we measured the qubit coherence time for a range of magnetic fluxes. The spin-echo dephasing rate is shown as a function of the flux sensitivity coefficient ${\partial \omega_{01}}/{\partial \Phi}$ in Fig.~\ref{fig:fig5}. The nearly linear dependence is in line with other experiments and indicative of low-frequency flux noise~\cite{yoshihara_decoherence_2006,orgiazzi_flux_2016}. Detailed measurements of coherence with CPMG pulse sequences were done at a flux $\Phi = 0.501\, \Phi_0$, for various pulse sequence lengths $N$. The dephasing rate changes from $1.4\,\upmu\t{s}$ for $N=1$ to $6.8\,\upmu\t{s}$ for $N=100$, with an approximately  $N^{0.42}$ dependence.  

To investigate multi-level control in the qutrit formed by the lowest three energy levels, we  characterize the properties of the readout in this space. First, we characterize the thermal state, described by the populations $P_\t{th0}$ and $P_\t{th1}$ for states 0 and 1 respectively. This is done based on Rabi oscillations after suitable swaps of states in the qutrit space~\cite{jin_thermal_2015,yan_distinguishing_2018}. The steady state populations correspond to an effective qubit temperature of 30~mK, very close to the cryostat temperature of 27 mK, in contrast to other reported results on long coherence time superconducting qubits, where larger differences between effective temperature and cryostat temperature were observed~\cite{reagor_quantum_2016, gustavsson_suppressing_2016}. Next, we characterize the homodyne voltages $V_\t{h0}$, $V_\t{h1}$, and $V_\t{h2}$ for the three qutrit states based on measuring the homodyne voltage as $V_\t{h} =P_0 V_\t{h0} + P_1 V_\t{h1} + P_2 V_\t{h2}$, where $P_0$, $P_1$, and $P_2$ are the probabilities of the qutrit states, with different initial preparations~\cite{yurtalan2020implementation}.
\begin{figure}[h]
\centering
\includegraphics[width=3.4in]{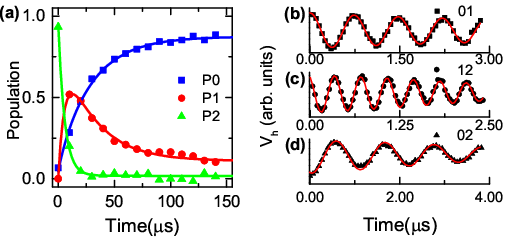}
\caption{\label{fig:3levels}(Color online) (a) The population versus the delay time after a $\pi_x^{01}$ and  a $\pi_x^{12}$ pulses are applied to the thermal state for levels 0 (blue squares), 1 (red dots), and 2 (green triangles). (b-d) Ramsey oscillations in time for coherent superposition of states (b) 0 and 1, (c) 1 and 2, and (d) 0 and 2, with Ramsey coherence times $4.7$, $3.4$, and $5.4$ $\upmu$s, respectively.}
\end{figure}
We performed experiments addressing coherence in the qutrit space. In Fig.~\ref{fig:3levels}(a), we show the result of a multi-level relaxation experiment. Two pulses, $\pi_x^{01}$ and $\pi_x^{12}$, are applied to excite the system to state $2$ and the populations are measured versus time. The relaxation dynamics  depends on the relaxation and excitation rates for each pair of states of the qutrit. The continuous lines are  fit with a multi-level relaxation model following Ref.~\cite{zizak_rate_1980}. In the fit, the 0-1 relaxation ($\Gamma_{10}$) and excitation ($\Gamma_{01}$) rates are set based on the measurements of the qubit relaxation time and thermal population described above. The ratios $\Gamma_{12}/\Gamma_{21}$ and $\Gamma_{02}/\Gamma_{20}$ are set equal to the corresponding Boltzmann factors, assuming the temperature extracted from qubit thermalization experiments, and $\Gamma_{21}$ and $\Gamma_{20}$ are free parameters. We note that the constraints on the ratios of the 1-2 and 0-2 rates do not significantly impact the values of the relaxation rates. Based on a fit of the variation of the populations with time, we extract the  relaxation and excitation rates at the flux symmetry point and away from the symmetry point, at $\Phi=0.501\,  \Phi_0$ as shown in Table~\ref{tbl:simrates}. 
\begin{table}[h]
	\centering
	\caption{Multi-level relaxation and excitation rates measured at the flux symmetry point  and away from the symmetry point.  }
	\begin{tabular*}{3.4in}{c@{\extracolsep{\fill}}c@{\extracolsep{\fill}}c}
		\hline
		\hline
		Rate & Value at 0.5 $\Phi_0$ & Value at 0.501 $\Phi_0$\\
		\hline
		$\Gamma_{01}$& 1.4 kHz & 1.2 kHz \\
		$\Gamma_{10}$& 29.5 kHz & 63.4 kHz \\
		$\Gamma_{12}$& 8.8 Hz & 0.4 Hz \\
		$\Gamma_{21}$& 124.3 kHz & 78.1 Hz \\
		$\Gamma_{02}$& 0.1  Hz & 0.01 Hz \\
		$\Gamma_{20}$& 27.8 kHz & 61.1 kHz \\
		\hline
		\hline
	\end{tabular*}
	\label{tbl:simrates}
\end{table}
  
 We also performed multi-level dephasing experiments for the three pairs of levels involved. Figure~\ref{fig:3levels}(b) shows Ramsey oscillations for coherent superpositions of states at the flux symmetry point. The decay curves are in excellent agreement with a model that includes state preparation, the interlevel relaxation and excitation rates, and low frequency noise modeled as a classical Gaussian stochastic process. The pure dephasing times extracted with this model and assuming a Gaussian decay are shown in the caption of Fig.~\ref{fig:3levels}. In addition, we characterized pure dephasing away from the symmetry point, at $\Phi = 0.501\, \Phi_0$. This experiment was done using a Ramsey protocol for superpositions of states 0-1 and 0-2, leading to  dephasing rates of $2.7$ MHz and $0.9$ MHz, respectively. 
 
 We next used the randomized benchmarking protocol~\cite{knill_randomized_2008,magesan_scalable_2011} to characterize the average fidelity of single-qubit gates. For the pulses corresponding to  the gates used in the sequences, having 1.62 ns and 2.64 ns duration for $\pi/2$ and $\pi$ rotations and a maximum drive strength of $2\pi\times260$ MHz,  the average  gate fidelity~\cite{magesan_scalable_2011} is found to be $99.92 \, \pm \, 0.003\%$.
\section{Discussion}
We now discuss the results obtained for coherence, starting with pure dephasing away from the circuit flux symmetry point. The nearly linear dependence of the Ramsey and spin-echo dephasing rates on the flux coupling coefficient ${\partial \omega_{01}}/{\partial \Phi}$ (see Fig.~\ref{fig:fig5}) demonstrates that dephasing away from the symmetry point is dominated by flux noise, and is indicative of flux noise with a PSD of the form $A/\abs{\omega}^\delta$, where $\omega$ is the frequency, $A$ characterizes the strength of the noise, and $\delta \approx 1$~\cite{yoshihara_decoherence_2006,orgiazzi_flux_2016}. At $\Phi = 0.501\, \Phi_0$, the coherence rate with CPMG pulses changes as $N^{-\beta}$. This is indicative of low-frequency noise with a power density $A/\abs{\omega}^\delta$, where $\delta = \beta / (1-\beta)$). This allows extracting $A = 1.8\times10^{-14}\, \t{(rad/s)}^{\delta-1}\Phi_0^2 $ and $\delta = 0.68$ (see Section~\ref{sec:cpmg2}); note that these values hold over the frequency range where the CPMG pulses are sensitive to flux noise, corresponding approximately to $3 - 46$~MHz. The Ramsey dephasing rates for 0-1 and 0-2 coherences at  $\Phi = 0.501\, \Phi_0$ are in a ratio proportional to the flux sensitivity coefficients, suggesting that flux noise is the dominant dephasing source for higher levels as well. 

Next we discuss the possible sources for the pure dephasing measured at the device flux symmetry point. We first consider the role of charge noise. We performed numerical simulations of the charge dispersion, yielding a charge modulation over one period of the island charges of $2\pi\times 133$ Hz, $2\pi\times 626$ Hz, and $2\pi\times 493$ Hz for the transitions 0-1, 1-2 and 0-2, respectively. The contribution of this source to the dephasing rate, which is bounded by the calculated modulation ranges, is therefore a negligible contribution to the measured rate. 

We next considered photon number fluctuations in the cavity. Based on numerical simulations of the spectrum of the coupled qubit/cavity system, we find that the dispersive shift for the $0-1$ transition is 0.5~MHz. To obtain this number we use the first several levels of the qubit and one level of the cavity. We note that in the experiments, we observed in certain cases, when the repetition time of experiments was too short, beating patterns in Ramsey oscillations with a beating frequency of $0.5$~MHz, which we attribute to photon number fluctuations. This value of the beating frequency is in  agreement with the numerically determined dispersive shift. Based on this value of the dispersive shift and the measured cavity bandwidth of 0.6 MHz, we estimate that a photon temperature of 250 mK is needed to induce a photon dephasing rate of 213 kHz, in line with the measured Ramsey coherence time; this temperature is unreasonably high. Next, we performed simulations of coherence decay for CPMG sequences. We found that a thermal photon number $n_\t{th} = 0.15$, corresponding to an effective temperature of 160 mK, is needed to explain experimental values for the CPMG decay times from one to ten CPMG pulses; this value still leads to decay times which are longer than measured for CPMG pulses with larger number of pulses. In addition, in a subsequent cooldown of the device, we performed spin-locking experiments using the method in Ref.~\cite{yan_distinguishing_2018}, which indicated an upper bound for the cavity temperature of 67 mK and a corresponding dephasing rate of 4.68 kHz. Based on these considerations, this source has a small contribution to the overall dephasing rate. 

Flux noise is another possible source of dephasing, even at the qubit symmetry point~\cite{makhlin_dephasing_2004}. We performed numerical simulations of the 0-1 transition frequency fluctuations arising from fluctuations of the magnetic flux. If we assume Gaussian flux noise with a PSD of $A/|\omega|^\delta$, with $A$ and $\delta$ based on two sets of experiments - the spin-echo versus flux and the CPMG versus the number of pulses at the $0.501\, \Phi_0$ flux bias point - the predicted dephasing times are significantly longer than experimentally measured. While this analysis indicates that flux noise is not a dominant source, we note that departures of the PSD from the simple form assumed could alter the predicted rates significantly. We note that experiments to measure the distribution of the transition frequencies showed a skewness, possibly indicative of the role of flux noise (see Appendix~\ref{sec:appflux}). 

For completeness, we mention that other possible sources of decoherence include two-level fluctuators (TLFs)~\cite{Klimov_2018_fluctuations} and the interaction with the second qubit on the chip. In a subsequent cooldown of the same device, the Ramsey dephasing time at the symmetry point increased to 10 $\upmu$s, while the energy gap and the persistent current of the device changed insignificantly, indicating a possible contribution to decoherence from fluctuating TLFs. While this analysis is not conclusive in terms of identifying the dominant dephasing mechanisms, future work will address this important aspect by exploring devices with a range of parameters affecting coupling to different noise sources.

We next discuss the observed energy relaxation times. We first consider the temporal fluctuations observed in qubit relaxation time. We performed analysis on a series of repeated 30 relaxation experiments based on a model that includes quasi-particle relaxation (see Appendix~\ref{sec:appt1}). We extracted the characteristic energy relaxation time $T_\t{1qp}$ due to the presence of a quasi-particle and the relaxation time $T_\t{1R}$ due to other noise channels. The analysis indicates $T_\t{1qp}=37.6\ \upmu$s and $T_\t{1R}=47.6\ \upmu$s. The fluctuations show an average of 0.23 $\pm$ 0.20  quasi-particles throughout the experiments, in agreement with the reported values in another work~\cite{yan_flux_2016}. Next, we investigated the possible sources for the measured $T_\t{1R}$. Estimates with $1/\abs{\omega}^\delta$ type flux noise show that flux noise could be  the dominant source limiting the relaxation time. Based on the extracted flux noise PSD by CPMG experiments and following the Fermi's golden rule~\cite{you_low-decoherence_2007}, we calculate the expected energy relaxation time to be in $24.7$ to $71.6\ \upmu$s range, where this range is due to the fit errors in determining the flux noise PSD. We note this calculation assumes the flux noise PSD being extrapolated to qubit frequencies, and may be affected by this approximation. Nevertheless, this estimate indicates that flux noise is likely the dominant contributor to energy relaxation at the symmetry point. We also measured the relaxation and excitation rates for the higher levels (see~\ref{tbl:simrates}). The decay rate for the 1-2 transition is higher than the 0-1 level, consistent with the larger transition strength. Due to the selection rules~\cite{liu_optical_2005}, at the symmetry point the 0-2 transition is forbidden for flux and charge coupled operators.  One possible source for observing a finite substantial rate for this transition is the  quasi-particle induced relaxation. 

We performed numerical simulations of the implemented randomized benchmarking experiment. In a three level simulation, we find that excitation to state 2 is negligible, in agreement with experiments where the population of state 2 is measured at the end of the random gate sequences (data not shown). Next, we performed simulations including only the two lowest (qubit) levels, with energy relaxation/excitation rates as measured and assuming pure dephasing with the Ramsey rate. These simulations give a fidelity of 99.95\% assuming the rotating wave approximation and a fidelity of 99.85\% if we include the counter rotating terms for the driving. This result indicates that counter rotating terms are likely the limitation in the current experiment, which can be improved on with pulse shaping~\cite{deng_observation_2015}. The lower fidelity in simulations is likely due to overestimating decoherence by assuming the Ramsey dephasing rate during driven evolution.  Our results, which do not yet use optimized control pulses, compare favorably with state-of-the-art numbers for other types of superconducting qubits. Transmons achieve gate fidelities of 99.9-$99.95\%$ for pulse durations in the 10-20 ns range~\cite{barends_superconducting_2014,sheldon2016characterizing,rol2017restless}. An experiment with pulse distortion compensation of flux qubits led to gate fidelities of $99.9\%$ with 4 ns pulses~\cite{gustavsson2013improving}. Non-adiabatic gates for fluxonium qubits reached $99.8\%$ fidelity with pulses in the 20-60 ns range~\cite{zhang2021universal}.

\section{Conclusions}
In conclusion, we implemented a new capacitively shunted flux qubit design that combines high anharmonicity, long relaxation  and  dephasing times.  The energy level structure and other properties are in excellent agreement with a model based on the numerically determined system capacitance matrix. We  demonstrated fast and  high fidelity single-qubit gates, characterized with randomized benchmarking. We expect that the use of optimal control will lead to improved single-qubit gates and to fast high fidelity two-qubit gates~\cite{zhu2021quantum}. The high anharmonicity of the circuit is  expected to be an advantage in quantum annealing applications~\cite{weber2017coherent,novikov2018exploring}. In addition, we demonstrated control and measured decoherence for the same circuit operated as a qutrit, based on the lowest three energy states. The measured decoherence is in good agreement with the multi-level decoherence model. Qutrits offer interesting perspectives for quantum error correction~\cite{prakash2018normal,anwar2014fast,krishna2019towards} and quantum simulation of topological states~\cite{haldane1983Continuum}. Future work will have to address the origin of dephasing and further design optimization to balance coherence and control speed.
%TC:ignore
\begin{acknowledgments}
We thank the University of Waterloo Quantum Nanofab team members for assistance on the device fabrication, Michal Kononenko for the help with experiments, and Sahel Ashhab for useful discussions on the experiments. We acknowledge support from the Natural Science and Engineering Research Council (NSERC), Canada Foundation for Innovation, Ontario Ministry of Research and Innovation, Industry Canada, and the Canadian Microelectronics Corporation. During part of  this work, A. L. was supported by an Early Research Award.
\end{acknowledgments}
%TC:endignore

\appendix
\appsection{Experimental details}
\subsection{Sample Fabrication}

The device is fabricated on a silicon substrate in two steps. In the first step, the CPW  resonator, the qubit control lines, and the junction shunt capacitors are defined by optical lithography. A $100$ nm thick aluminum layer is deposited by e-beam evaporation, which is then followed by a lift-off process. In the second step, the Josephson junctions are first patterned on  a bi-layer e-beam resist stack (PMGI SF7 + PMMA 950K A3) using e-beam lithography. The junctions are made with  double-angle shadow evaporation of two aluminum layers of thicknesses $40$ nm and $65$ nm respectively, with an oxidation carried out after the deposition of the first layer. Argon milling  prior to the evaporation of the two aluminum layers is used to ensure a high-quality contact between the junctions and the aluminum layer evaporated in the first step.

\subsection{Experimental setup}

The experiments are performed in a dilution refrigerator with a base temperature of 27~mK. The sample is enclosed in a copper box. Superconducting coils attached to this box are used for magnetic flux biasing. To implement magnetic shielding,  the sample box is placed inside a three layer high-magnetic-permeability metal shield. The sample is connected to readout and control equipment  with coaxial cables and the signals are attenuated and filtered at different temperature stages of the refrigerator.

Qubit control pulses are generated using two different configurations. In the first configuration, pulses are directly generated  by a Tektronix AWG70001A arbitrary waveform generator (AWG). The AWG has a 50 GS/s sampling rate and   10 bit amplitude resolution. Most experiments are performed using this setup. In the second configuration, the pulses are generated by standard microwave synthesizers and shaped with IQ mixers using signals generated by a lower bandwidth Tektronix 5014 AWG. This second configuration was used primarily for preliminary device characterization. For readout, we use microwave pulses generated using a synthesizer and mixers. After transmission through the cavity, pulses are amplified using a cryogenic HEMT amplifier and room temperature amplifiers; after demodulation, the pulses are recorded with a digitizer. 

\appsection{Design Analysis}
\label{sec:appopt}
In this section we discuss a numerics based approach for optimization of a capacitively shunted flux qubit. We assume a general design with three Josephson junctions, out of which two are equal in size and the third one is smaller by a factor  $\alpha$. This configuration results in a highly anharmonic potential when the loop is biased at half a flux quantum. The potential is a double well potential for $\alpha>0.5$~\cite{orlando_superconducting_1999,mooij_Josephson_1999} and it still retains useful properties for $\alpha \le 0.5$. In addition, we assume that there are large added capacitances shunting the junctions and to the ground; the capacitance matrix is symmetric, as the junctions. These assumptions, together with given junction critical current density ${J}_\t{c}$, and junction capacitance density $\tilde{C}$, leave six parameters for design: 

\begin{itemize}
    \item $\lambda_1$ - the geometric capacitance between pads 2 and 3, 
    \item $\lambda_2$ - the area of the smaller junction, 
    \item $\lambda_3$ - the ratio of the area of the smaller junction to the area of the large junctions,
    \item $\lambda_4$ - the ratio of the geometric capacitance between pads 1 and 2 (or equivalently 1 and 3) to the geometric capacitance between pads 2 and 3,
    \item $\lambda_5$ - the ratio of geometric capacitance between pad 1 and ground to the geometric capacitance between pads 2 and 3, and
    \item $\lambda_6$, the ratio of the geometric capacitance between pad 2 (or equivalently pad 3) and ground to the geometric capacitance between pads 2 and 3.
\end{itemize}
The device metrics are then calculated with the circuit model in Eq.~\ref{eq:ham} with the parameters given in Table~\ref{tbl:simopt}.
\begin{table}[h]
	\centering
	\caption{ The device design optimization parameters (see text for explanations). }
	\begin{tabular*}{3.4in}{@{\extracolsep{\fill}}rclp{0.5cm}rcl}
		\hline
		\hline
		$C_{23}$&=&$\lambda_1 + \lambda_2 \tilde{C}$  &&
		$C_{12}, C_{13}$&=&$ \lambda_1 \lambda_4 + \lambda_2 \tilde{C} / \lambda_3 $  \\
		$C_{01} $&=&$ \lambda_{1} \lambda_5$  &&
        $C_{02}, C_{03} $&=&$ \lambda_{1} \lambda_6$  \\
		$I_\t{c}$ &=& $J_\t{c} \lambda_{2}/\lambda_{3}$  &&
		$\alpha_\t{J}$&=&$\lambda_{3}$  \\
		\hline
		\hline
	\end{tabular*}
	\label{tbl:simopt}
\end{table}

The optimization is designed to find target values for the following parameters: the transition frequency $\omega_{01}$ and anharmonicity at the flux symmetry point, the persistent current, and the offset charge modulation. Their relevance is as follows. The transition frequency determines the qubit thermal state, which is ideally larger than the temperature; note however that alternative state preparation methods can be used to prepare the qubit in a pure state~\cite{LupascuHigh-Contrast2006,Valenzuela1589,geerlings_demonstrating_2013}. The transition frequency and the anharmonicity determine the speed of the gates, as limited by the rotating wave approximation and leakage to other states respectively. The persistent current in combination with $\omega_{01}$ determine the dephasing at the symmetry point. Finally, the modulation with offset charges determines decoherence due to charge noise and quasi-particle induced transitions. 

This framework allows for a comparison of the flexibility enabled by the three pad design explored in this work versus  two pad designs~\cite{yan_flux_2016,steffen_high-coherence_2010}. We applied this optimization with the parameters of the device in this work as a target and with an assumed charge modulation below 1 kHz, in line with the observed coherence times. The simulation converges to values that are close to our design parameters and in agreement with the refined parameters from the fit of the spectroscopy data. In addition, we attempted to reach the same qubit parameters with a two-pad design. No design parameters were found to reach these metrics; it was possible for example to reach the same transition frequency,  persistent current, and anharmonicity within 95$\%$ accuracy at the cost of larger charge modulation ( of the order of  MHz ).

\appsection{Population Extraction and Qubit Effective temperature}

In this section, we discuss the characterization of the thermal state population of the qubit. It is assumed that the population of state 2 is negligible as the transition frequency from the ground state to this state at the symmetry point, $\omega_{02}=2\pi\times 7.107$ GHz, is large compared to the temperature. The thermal state populations $P_{\t{th0}}$ and $P_{\t{th1}}$ of state 0 and 1 respectively are determined by performing Rabi oscillations on the 0-1 transition, with two different starting states. The first state preparation is obtained by waiting long enough for the system to thermalize and then applying  a $\pi_x^{12}$ pulse. The second state preparation is done by applying a $\pi_x^{01}$ pulse followed by a $\pi_x^{12}$ pulse to the thermalized state. This method is similar to the approach used in Jin \emph{et al}.~\cite{jin_ThermalResidualExcitedState_2015}, with the difference that the preparations are chosen such that Rabi oscillations for the 0-1 transition are compared, where the readout contrast is optimized for this qubit. The thermal state population $P_\t{th0}$ can be found as
\be
P_\t{th0}= \frac{A_0}{A_1+A_0},
\ee
where $A_0$ and $A_1$ are Rabi oscillation amplitudes corresponding to the two preparations described above. In the course of the experiments  $P_\t{th0}$ is found to be $0.95\pm 0.02$ indicating an effective temperature, calculated based on $P_\t{th1}/P_\t{th0}=\exp(-\hbar\omega_{01}/K_\t{B} T)$, of $27$-$32$ mK.

\appsection{Fluctuations in T1 relaxation time}
\label{sec:appt1}
In the course of the experiments, T1 relaxation time is observed to be fluctuating in time. This behavior is attributed to the presence of fluctuating excess quasi-particles and the excited state population decay  is expressed by \cite{pop2014coherent} 
\begin{equation}
   P(t)=e^{\lambda_\t{qp}[ \exp{(-t/T_\t{1qp})-1}]}e^{-t/T_\t{1R}},
    \label{eq:qp1}
\end{equation}
where $\lambda_\t{qp}$ is the mean-value of the Poisson probability distribution of quasi-particles, $T_\t{1qp}$ is the relaxation time in the presence of 1 quasi-particle, and $T_\t{1R}$ is the relaxation time due to other loss channels in the absence of quasi-particles. We  repeatedly measured the T1 relaxation times after exciting the thermal state with  $(\pi)_x^{01}$ pulse   and fitted the decay curves with Eq.~\ref{eq:qp1}, with $T_\t{1qp}$ and $T_\t{1R}$ being global fit parameters for all decay curves. 
\begin{figure}[h]
	\centering
	\includegraphics[width=3.4in]{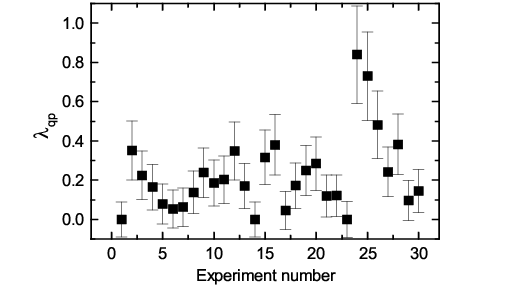}
	\caption{\label{fig:nqp} The number of quasi-particles in repeated relaxation experiments.}
\end{figure}
 Figure~\ref{fig:nqp} shows the fluctuations in the number of quasi-particles in subsequently measured 30 relaxation times with the extracted $T_\t{1qp} = 37.6$ $\upmu$s and $T_\t{1R} = 47.6$ $\upmu$s, where the experiment repetition period is 9 minutes. We note a mean of 0.23 quasi-particles throughout the experiments with a standard deviation of 0.20.

\appsection{Flux noise simulations}
\label{sec:appflux}
The measurements of coherence away from the symmetry point are consistent with flux noise with a $A/\abs{\omega}^\delta$ power spectral density. We discuss decoherence induced by flux noise at the qubit symmetry point, where flux fluctuations are coupled quadratically to flux. The quadratic coupling does not allow finding a simple expression for the coherence function, as it is possible for linearly coupled noise (see Eq.~\ref{eq:CPMGCoherenceDecay}).

Analytical expressions in limit cases for decoherence with quadratic coupling for white and $A/\abs{\omega}$ noise were derived by Makhlin and Shnirman~\cite{makhlin_dephasing_2004}. Here, we resort to a numerical simulation of the experiment. We numerically generate flux noise trajectories as follows. For a coherence probing experiment of duration $\tau$, we generated multiple random noise samples. Each noise sample consists of $N_s$ random values $\xi_i = \xi (\tau / N_s  (i-1))$, $i = \overline{1,N_s}$. The sample ${\vect\xi} = ( \xi_1, \xi_2, \xi_3, .., \xi_{N_s} )$ is drawn from the multivariable Gaussian distribution $\exp{\left(-\frac{1}{2}{\vect\xi}^T {\mathbf{M}}^{-1}{\vect\xi}\right)}$, where ${\mathbf{M}}$ is the matrix of correlation coefficients. The correlation coefficients M$_{ij}$ are obtained by numerical integration.

We verified our numerical method by doing a numerical simulation of a CPMG experiment away from the symmetry point. Specifically, we consider CPMG sequences with $N= 1, 5, 10, 20, 40,\t{ and }100$. For each $N$ value, we consider a set of evolution times $\tau$ ranging from $1$ to $20\,  \upmu$s. For each $\tau$ we generate $1024$ noise samples and obtain the coherence time as an average of the off-diagonal part of the density matrix over the noise samples. Using the noise PSD of $1.8\times 10^{-14} /|\omega|^\delta (\t{rad/s})^{\delta-1}\Phi_0^2$ with $\delta=0.68$, determined from experimental data, the CPMG decay curves from numerical simulations are in excellent agreement with experimental results and Eq.~\ref{eq:coh3}.

\begin{figure}[h!]
	\centering
	\includegraphics[width=3.2in]{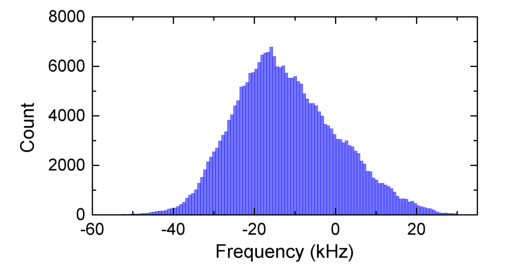}
	\caption{\label{fig:histfreq} The histogram of the qubit frequency fluctuations at flux symmetry point.  }
\end{figure}

\appsection{Experimentally observed frequency fluctuations}
In a subsequent cooldown of the device, we measured the frequency fluctuations of the qubit in a protocol similar to Ramsey experiments. A coherent superposition of  states  0 and 1  is prepared by applying a $(\pi/2)_x^{01}$ pulse, with  the drive frequency slightly detuned from the 0-1 transition frequency $\omega_{01}$. Next, after waiting for some fixed evolution time, in which the detuning and qubit frequency fluctuations are reflected in the  acquired phase,  another $(\pi/2)_x^{01}$ pulse is applied and the state readout is performed. In order to map the readout homodyne voltage to detuning in frequency, in a similar experiment, the drive frequency (or detuning) is swept and readout homodyne voltage is recorded. With this, the fluctuations in the readout homodyne voltage are converted to frequency fluctuations.  This protocol is repeated  $2^{18}$ times with 200 $\upmu$s  repetition time. The histogram of the frequency fluctuations around qubit nominal frequency, shown in Fig.~\ref{fig:histfreq}, shows a significant skewness and is indicative of flux noise being a possible significant contributor to dephasing.

\appsection{Randomized benchmarking}
The pulses in randomized benchmarking experiments, corresponding to the unitary operators from the Clifford group and the Pauli group, are   $\pi/2$ and $\pi$ rotations around the x(y) axis and denoted by $R_{x(y)}(\theta)$ where $\theta$ is the rotation angle. The calibration procedure for the $\pi/2$ pulses is done by applying $ [R_{x(y)}(\pi / 2)]^{2m+1}$ pulses with $m$ an integer. For large number of repetitions the accumulated error is projected on the measurement basis and the error is minimized by adjustments on $\theta$.  Similarly, $\pi$ rotations are calibrated by applying $R_{x(y)}(\pi/2)[R_{x(y)}(\pi)]^{(2m+1)}$.
\begin{figure}[h!]
	\centering
	\includegraphics[width=3.1in]{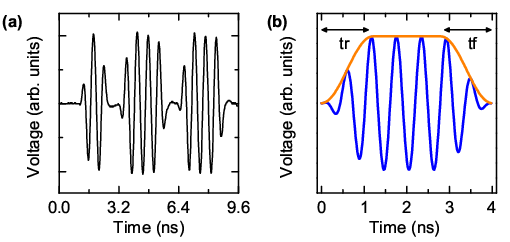}
	\caption{\label{fig:rbpulse}(a) A section of the randomized benchmarking pulse sequence measured  with an oscilloscope. (b) An example software generated single shaped pulse showing the rise time $\t{t}_\t{r}$ and the  fall time $\t{t}_\t{f}$.  }
\end{figure}
A section of  the waveform of the  sequence pulse used in the experiments  is measured with an oscilloscope and  shown in Fig.~\ref{fig:rbpulse}. The pulses have a driving strength of $2\pi\times260$ MHz and rise and fall times of $0.6$ ns. The duration of the $\pi/2$ and $\pi$ pulses are $1.62$ and $2.64$ ns respectively. The time gap between each pulse in the sequence is $0.5$ ns. Oscilloscope measurements  indicate a good control of pulse shaping with the Tektronix AWG70001A. 
\begin{figure}[hbt!]
\includegraphics[width=3.4in]{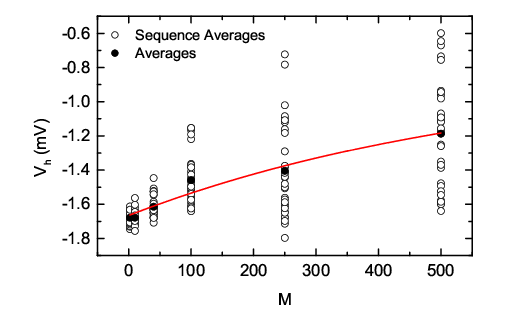}
\caption{The homodyne voltage  after randomized benchmarking sequence is applied to qubit thermal state versus the sequence length $N$. The black circles represent the measured voltage for each randomization of the sequence and the black dots represent the sequence averages. The solid curve represents the fit with the  decay function.  }
\label{fig:fig6}
\end{figure}

Figure~\ref{fig:fig6} shows the average measured homodyne voltage after evolution by each randomized sequence (black circles) versus the sequence length $M$.  The average of 32 sequences at each length N are indicated by black dots. The average of sequences has a decaying behaviour and  is fitted with  a function $F=A_0 p^M+B_0$ where $A_0$ and $B_0$ are fit parameters corresponding to the errors associated with state preparation and  measurement and $p$ is related to the average error rate~\cite{magesan_scalable_2011}. The average  gate fidelity $F_\t{ave}=p+(1-p)/2$ is found to be $99.92\pm0.003\%$.

%REFERENCES

%When using a bibtex file (the only option you should consider at the draft stage), use the section below.
%\bibliography{AnharmonicCSFQbib}%bibtex file, use path relative to the folder

%

\end{document}